\documentclass[prd,showpacs,superscriptaddress,nofootinbib,floatfix,11pt]{revtex4-2}
\usepackage[utf8]{inputenc}
\usepackage{amsmath,amssymb}
\usepackage{slashed}
\usepackage{subfigure}
\usepackage[colorlinks=true,
breaklinks=true,
urlcolor=magenta,
citecolor=blue]{hyperref}
\usepackage[usenames,dvipsnames]{color}
\usepackage{appendix}
\usepackage{braket,bm}
\usepackage{multirow}
\usepackage{enumitem}
\usepackage{color}
\usepackage{cancel}
\usepackage{dsfont}

\usepackage{graphicx}
\usepackage{tikz}
\usepackage{booktabs}
\usepackage{array}
\usepackage{overpic}
\usepackage{float}
\usepackage{threeparttable}
\allowdisplaybreaks[4]

\usetikzlibrary{calc}
\usetikzlibrary{intersections}
\usetikzlibrary{trees}
\usetikzlibrary{decorations.pathmorphing}
\usetikzlibrary{decorations.markings}
\usetikzlibrary{arrows.meta}
\usetikzlibrary{patterns}
\tikzset{
   global scale/.style={
      scale=#1,
      every node/.append style={scale=#1}},
   photon/.style={decorate, decoration={snake}, draw=red},
   nucleon/.style={draw=black, postaction={decorate},
      decoration={markings,mark=at position .65 with{\arrow[draw=black]{latex}}}},
   pion/.style={draw=blue, postaction={decorate},
      decoration={markings,mark=at position .55 with{\arrow[draw=blue]{}}}},
    nucleonstar/.style={draw=black, postaction={decorate},
      decoration={markings, mark=at position 0.7 with {\arrow[draw=black]{latex}}}},
    }

\newcommand{\itp}
{\affiliation{CAS Key Laboratory of Theoretical Physics, Institute of Theoretical Physics, Chinese Academy of Sciences, Beijing 100190, China}}

\newcommand{\hebtu}{\affiliation{Department of Physics and Hebei Key Laboratory of Photophysics Research and Application,\\
Hebei Normal University, Shijiazhuang 050024, China}}

\newcommand{\md}{\mathrm{d}}

\begin{document}
\title{Comprehensive study of axion photoproduction off the nucleon in chiral effective field theory} 

\author{Xiong-Hui Cao}\email{xhcao@itp.ac.cn}
\itp
\author{Zhi-Hui Guo}\email{zhguo@hebtu.edu.cn}
\hebtu

\begin{abstract}
We calculate the amplitudes of the axion photoproduction off the nucleon, i.e., $\gamma N \to a N$, within the framework of chiral effective field theory. Several different types of contributions are simultaneously included in our calculation, namely the nucleon exchanges up to next-to-leading order, the $a\gamma\gamma$ vertex and the vector-meson exchanges in the $t$ channel. We utilize the existing hadronic inputs as much as possible to fix the unknown couplings. A comprehensive study of the phenomenological discussions is then provided in this work. Different mechanisms in the $\gamma N \to a N$ processes manifest distinct behaviors in the total and differential cross sections, which could provide useful quantities to distinguish different axion models. 
\end{abstract}

\maketitle

\section{Introduction}

The hunt of the axion imprint constitutes one of the most active research subjects in particle physics, cosmology and even atomic and molecular physics~\cite{Kim:2008hd,Graham:2015ouw,Irastorza:2018dyq,Sikivie:2020zpn,Choi:2020rgn,DiLuzio:2020wdo}. Among the plethora of proposed hunting methods, the axion photoproduction off a nucleon target offers an invaluable environment to simultaneously explore the axion parameters in different experimental contexts, such as the electron beam dump facility~\cite{Dusaev:2020gxi}, reactor-based experiment~\cite{AristizabalSierra:2020rom}, astrophysical objects including supernova and neutron stars~\cite{Lucente:2022vuo,Chakraborty:2024tyx,Dev:2023hax}, electron-ion colliders~\cite{Pybus:2023yex,Anderle:2021wcy},  etc. 

The amplitudes of the axion photoproduction processes, i.e., $\gamma N \to a N$, are the key inputs to bridge the experimental constraints and the underlying theory. Nowadays many works often rely on simple phenomenological models to parametrize such amplitudes~\cite{AristizabalSierra:2020rom,Dusaev:2020gxi,Lucente:2022vuo,Chakraborty:2024tyx,Dev:2023hax,Pybus:2023yex}, which usually only include parts of some leading-order contributions, such as those from the $a\gamma\gamma$ vertex. The primary purpose of this work is to pursue a more systematical calculation of the axion photoproduction amplitudes within the framework of chiral effective field theory (EFT). We aim to carry out a more complete calculation of the $\gamma N \to a N$ processes below the $\pi N$ threshold by simultaneously including the nucleon exchanges in the $s$ and $u$ channels, the contribution from the $a\gamma\gamma$ vertex, and the vector resonance exchanges in the $t$ channel. The hadronic inputs from various processes in literature will be greatly exploited to fix the unknown couplings, which allows us to discern the relative importance of different pieces and to distinguish possible signatures from different mechanisms. We foresee that the improved axion photoproduction amplitudes would provide useful inputs for future theoretical studies in other contexts, such as the astrophysical and cosmological environments, and also experimental analyses. 

This paper is structured as follows. In Sec.~\ref{sec.achpt}, we elaborate the pertinent axion chiral Lagrangians with baryons and calculate the nucleon-exchange parts of the $\gamma N\to a N$ amplitudes. The anomalous vertices of the $a\gamma\gamma$ and $Va\gamma$ types are discussed, and their contributions to the $\gamma N\to a N$ processes are calculated in Sec.~\ref{sec.agg}. We then present the phenomenological discussions by separately studying different kinds of contributions to the total and differential cross sections in Sec.~\ref{sec.phenon}. A short summary and conclusions are given in Sec.~\ref{sec.summary}.

\section{Relevant axion interactions in chiral EFT}\label{sec.achpt}

The axion mechanism is originally proposed by Peccei and Quinn~\cite{Peccei:1977hh,Peccei:1977ur} to solve the strong $CP$ problem, and the spontaneous breaking of the global Peccei-Quinn (PQ) $U(1)$ symmetry and the anomalous axion-gluon interaction render the axion to be a pseudo-Nambu-Goldstone boson (pNGB)~\cite{Peccei:1977hh,Peccei:1977ur,Weinberg:1977ma,Wilczek:1977pj}. 
The general QCD Lagrangian including light quarks, gluons and the axion/axion-like particle field (in the following we simply denote them as axion) $a(x)$ reads 
\begin{align}\label{eq.L1}
    \mathcal{L}_{\mathrm{QCD}}^{\mathrm{axion}}=\mathcal{L}_{\mathrm{QCD},0}-\bar{q} \mathcal{M}_q q+\frac{1}{2}\partial_\mu a\partial^\mu a-\frac{1}{2}m^2_{a,0}a^2 +\frac{a}{f_a} \frac{g_s^2}{16 \pi^2} \operatorname{Tr}\left[G_{\mu \nu} \tilde{G}^{\mu \nu}\right]+\frac{\partial^\mu a}{2 f_a} J_\mu^{\mathrm{PQ}}\ ,
\end{align}
where $\mathcal{L}_{\mathrm{QCD},0}$ denotes the conventional QCD Lagrangian excluding the quark mass terms and the light-flavor quark contents with $q=(u, d, s)^{T}$ will be focused in this work. $f_a$ stands for the axion decay constant, and $m_{a,0}$ is a bare PQ symmetry breaking contribution to the axion mass (for QCD axion $m_{a,0}=0$). The quark mass matrix is already in the diagonal real basis $\mathcal{M}_q=\operatorname{diag}\left(m_u, m_d, m_s\right)$, and $g_s$ is the QCD strong interaction coupling constant. $G_{\mu \nu}=G_{\mu \nu}^a \lambda^a / 2$ and $\tilde{G}_{\mu \nu}=\frac{1}{2} \epsilon_{\mu \nu \alpha \beta} G^{\alpha \beta}$ correspond to the gluon field strength tensor and its dual\footnote{We use the convention $\epsilon_{0123}=+1$ for the Levi-Civita tensor $\epsilon_{\mu \nu \alpha \beta}$ throughout this work.}, with $\lambda^a$ the Gell-Mann matrices. 

The anomalous $a G\tilde{G}$ term in Eq.~\eqref{eq.L1} is considered to be the model-independent part in diverse axion model constructions, while the derivative axion-quark interaction, i.e. the last term of Eq.~\eqref{eq.L1}, is variant in different axion models. The preexisting axial-vector quark current, sometimes also referred as the PQ current, is usually parametrized as 
\begin{align}
J_\mu^{\mathrm{PQ}}= \bar{q} \gamma_\mu \gamma_5 \mathcal{X}_q q\ ,
\end{align}
where  $\mathcal{X}_q=\operatorname{diag}\left(X_u,X_d,X_s\right)$ denotes the model-dependent coupling constant matrix in the flavor space. For example, the matrix elements take the forms
\begin{align}
    \begin{aligned}
    X_q^{\mathrm{KSVZ}} & =0\ , \\
    X_{u}^{\mathrm{DFSZ}} &=\frac{1}{3} \sin ^2 \beta\ , \quad 
    X_{d, s}^{\mathrm{DFSZ}} &=\frac{1}{3} \cos ^2 \beta\ ,
    \end{aligned}
\end{align}
in the Kim-Shifman-Vainshtein-Zakharov (KSVZ)~\cite{Kim:1979if,Shifman:1979if} and Dine-Fischler-Srednicki-Zhitnitsky (DFSZ)~\cite{Dine:1981rt,Zhitnitsky:1980tq} axion models, respectively, with $\tan\beta=v_u/v_d$ the ratio of the vacuum expectation values of the two Higgs doublets. 

It is customary to perform an axial rotation on the quark fields in order to remove the term $\propto a \operatorname{Tr}\left[G_{\mu \nu} \tilde{G}^{\mu \nu}\right]$ in Eq.~\eqref{eq.L1} by transforming $q \rightarrow \exp \left(i \gamma_5 \frac{a}{2 f_a} \mathcal{Q}_a\right) q $ with 
\begin{align}\label{eq.qa}
\mathcal{Q}_a=\frac{\mathcal{M}_q^{-1}}{\operatorname{Tr} \mathcal{M}_q^{-1}} = \frac{1}{1+z+w} \operatorname{diag}(1, z, w) \,, 
 \qquad \left(z=\frac{m_u}{m_d}, \, w=\frac{m_u}{m_s} \right) \ .
\end{align}
The particular form of $\mathcal{Q}_a$ has been chosen in order to avoid the leading-order meson-axion mass mixing~\cite{Georgi:1986df}. After this transformation, the Lagrangian in Eq.~\eqref{eq.L1} becomes
\begin{align}\label{eq.L2}
    \mathcal{L}_{\mathrm{QCD}}^{\mathrm{axion},\prime}=\mathcal{L}_{\mathrm{QCD},0}+\frac{1}{2}\partial_\mu a\partial^\mu a-\frac{1}{2}m^2_{a,0}a^2 -\bar{q} \mathcal{M}_a q +\frac{\partial^\mu a}{2 f_a} J_\mu^a \ ,
\end{align}
where the nonderivative axion-quark interactions are now entirely shifted into the phase of the quark mass matrix
\begin{align}\label{eq.Ma}
    \mathcal{M}_a=\exp \left(-i \frac{a}{2f_a} \mathcal{Q}_a\right) \mathcal{M}_q \exp \left(-i \frac{a}{2f_a} \mathcal{Q}_a\right) \ .
\end{align}
Because of the aforementioned quark transformation, the derivative axion-quark interactions get modified and an additional term appears in the quark axial-vector current 
\begin{align}\label{eq.current1}
    J_\mu^a=J_\mu^{\mathrm{PQ}}-\bar{q} \gamma_\mu \gamma_5 \mathcal{Q}_a q\ .
\end{align}
It is convenient to split $J_\mu^a$ into the isoscalar and isovector pieces for later chiral EFT constructions. The decomposition of the axial-vector currents~\eqref{eq.current1} into the $u/d$ and $s$ quarks can be written as 
\begin{align}\label{eq.axialc}
    J_\mu^{a, u d} & = \big(c_{u-d} \bar{q}  \gamma_\mu \gamma_5 \tau^3 q  +c_{u+d} \bar{q}  \gamma_\mu \gamma_5 q \big)_{q={(u,d)^{T}}}\ , \qquad 
    J_\mu^{a, s}  = c_{s} \bar{s}  \gamma_\mu \gamma_5 s\,,  
\end{align}
where we introduce the abbreviations $c_{u \pm d}$ and $c_s$ as~\cite{Vonk:2020zfh}
\begin{align}
    c_{u \pm d}=\frac{1}{2}\left(X_u \pm X_d-\frac{1 \pm z}{1+z+w}\right) \ , 
    \qquad c_s=X_s-\frac{w}{1+z+w}  \ .
\end{align}
Now the general axion-quark interaction Lagrangian from Eq.~\eqref{eq.L2} takes the form 
\begin{align}\label{eq.L3}
    \mathcal{L}_{a-q}=  -\bar{q}  \mathcal{M}_a q 
     +\frac{\partial_\mu a}{2 f_a} \big[\bar{q} \gamma^\mu \left(c_{u-d} \tau^3+c_{u+d} \mathds{1}\right) \gamma_5 q \big]_{q=(u, d)^{T}} 
    + c_s \bar{s} \gamma^\mu \frac{\partial_\mu a}{2 f_a} \mathds{1} \gamma_5 s\,,
\end{align}
which is now given in a proper basis to match the isovector and isoscalar parts of the
axial-vector currents to chiral EFT. 

A convenient way to match the axion-quark interactions in Eq.~\eqref{eq.L3} to the axion-hadron counterparts in chiral EFT is to use the external source method proposed in Refs.~\cite{Gasser:1983yg,Gasser:1984gg}. 
The axial external sources are taken to be traceless in order to avoid subtleties arising from the $U(1)_A$ anomaly in the former two references. However, the isoscalar or isosinglet axial current that has nonvanishing trace appears in the Lagrangian~\eqref{eq.L3}. To account for this isosinglet axial current in chiral EFT, we simply introduce an additional isosinglet axial external source, in analogy to the traceless axial currents. Of course, the coupling accompanied with the isosinglet axial current differs from that of the isotriplet one. According to the Lagrangian~\eqref{eq.L3}, one can define the following axial external sources to include the axion field  
\begin{align}\label{eq.asource}
     a_\mu^{(3)}=c_{u-d} \frac{\partial_\mu a}{2 f_a} \tau^3, \quad a_{\mu, i}^{(0)}=c_{i} \frac{\partial_\mu a}{2 f_a} \mathds{1}\ , \quad (i=u+d,s) 
\end{align}
for the isotriplet and isosinglet currents, respectively. 

Since we focus on the axion photoproduction off the nucleon target, i.e., the dynamical degrees of freedom for baryons are proton and neutron only, the $SU(2)$ version of chiral EFT with baryons will be employed to calculate the $\gamma N\to a N$ amplitudes, although the isosinglet external currents with both $u/d$ and $s$ quarks will be included. The usual $SU(2)$ matrix containing the three pions can be parametrized as
\begin{align}
u=\sqrt{U}=\exp \left(i \frac{\pi^a \tau^a}{2 F}\right)\ ,
\end{align}
where $\tau^a$ are the Pauli matrices, $F$ is the pion decay constant in the chiral limit, with the normalization of its physical value $F_\pi=92.1~\mathrm{MeV}$~\cite{ParticleDataGroup:2022pth}. It is noted that the difference between $F$ and $F_\pi$ amounts to the effects of higher orders than those considered here. 

The chiral connection and chiral vielbein with external sources are given by~\cite{Gasser:1983yg,Gasser:1984gg}  
\begin{align}
 \Gamma_\mu= \frac{1}{2}\left(u^{\dagger} \partial_\mu u+u \partial_\mu u^{\dagger}-i u^{\dagger} v_\mu u-i u v_\mu u^{\dagger}-i u^{\dagger} a_\mu u+i u a_\mu u^{\dagger}\right) \ , 
\end{align}
\begin{align}
    u_\mu=i\left(u^{\dagger} \partial_\mu u-u \partial_\mu u^{\dagger}-i u^{\dagger} v_\mu u+i u v_\mu u^{\dagger}-i u^{\dagger} a_\mu u-i u a_\mu u^{\dagger}\right)\ ,
\end{align}
respectively, where the relevant parts to the present calculations are obtained by taking the axial-vector external sources in Eq.~\eqref{eq.asource}, i.e. $a_\mu=a_\mu^{(3)}$. While, for the isoscalar sources $a_{\mu, i=(u+d,s)}^{(0)}$ in Eq.~\eqref{eq.asource}, we can introduce similar objects~\cite{Vonk:2020zfh}
\begin{align}
    \tilde{\Gamma}_\mu & =\frac{1}{2}\left(-i u^{\dagger} a_{\mu, i}^{(0)} u+i u a_{\mu, i}^{(0)} u^{\dagger}\right)\ ,\\
    \tilde{u}_{\mu, i} & =i\left(-i u^{\dagger} a_{\mu, i}^{(0)} u-i u a_{\mu, i}^{(0)} u^{\dagger}\right)\ .
\end{align}
It is easy to verify that the isoscalar components $a^{(0)}_{\mu, i=(u+d,s)}$ vanish in  $\tilde{\Gamma}_\mu$. The vector external source $v_\mu$ corresponds to the photon field in our study, and in the $SU(2)$ chiral EFT is given by  
\begin{align}\label{eq.vsource}
    v_\mu^{(\gamma)}=eQ_N A_\mu\ ,\quad Q_N=\frac{1}{2}\left(\mathds{1}+\tau^3\right)\ .
\end{align}
The field strength tensor of the photon field reads 
\begin{align}
    F^\pm_{\mu\nu}=u^\dagger f_{\mu\nu}u\pm uf_{\mu\nu}u^\dagger\ ,\quad f_{\mu\nu}=eQ_N\left(\partial_\mu A_\nu-\partial_\nu A_\mu\right)\ . 
\end{align}
For the nonderivative axion-hadron interaction, it can be embedded in the axion dressed quark mass term via the following building blocks
\begin{align}
    \chi_{a \pm}=u^{\dagger} \chi_a u^{\dagger} \pm u \chi_a^{\dagger} u\ ,
\end{align}
where $\chi_a=2 B\mathcal{M}_a$ with $\mathcal{M}_a$ in Eq.~\eqref{eq.Ma} and $B$ is a constant related to quark condensate via $B=\lim _{m_u, m_d \rightarrow 0}\left(-\langle\bar{u} u\rangle / F^2\right)$. 
It is noted that we closely follow the standard convention in Refs.~\cite{Gasser:1983yg,Gasser:1984gg,Scherer:2012xha} for the chiral transformation behaviors of the various building blocks, so that the sign convention of Eq.~\eqref{eq.Ma} is consistent with the sign of the axial-vector current in Eq.~\eqref{eq.axialc}.~\footnote{For example, there is a minus sign difference between the definition of the axion dressed quark mass in Ref.~\cite{Vonk:2020zfh} and ours in Eq.~\eqref{eq.Ma}.}

Collecting the proton and neutron fields in the isodoublet $N(x)=(p, n)^{T}$, the $SU(2)$ baryon chiral Lagrangian with axion up to $\mathcal{O}\left(p^2\right)$ in the low-energy expansion reads 
\begin{align}
    \mathcal{L}_{a N}=\mathcal{L}_{a N}^{(1)}+\mathcal{L}_{a N}^{(2)}+\ldots\ ,
\end{align}
with 
\begin{align}\label{eq.lagan1}
    \mathcal{L}_{a N}^{(1)}&=\bar{N}\left(i \slashed{D}-m_N+\frac{g_A}{2} \slashed{u} \gamma_5+\frac{g_{0}^{i}}{2} \slashed{\tilde{u}}_{i} \gamma_5\right)N\ ,\\
    \mathcal{L}_{a N}^{(2)}&=\bar{N}\left(\frac{c_6}{8 m_N} F_{\mu \nu}^{+}+\frac{c_7}{8 m_N} \operatorname{Tr}\left[F_{\mu \nu}^{+}\right]\right)\sigma^{\mu \nu} N+\cdots\ , \label{eq.lagan2}
\end{align}
where $D_\mu=\partial_\mu+\Gamma_\mu$ is the covariant derivative. $m_N$ is the leading-order (LO) nucleon mass, and $g_A$ and $g_{0}^{i}$ are the LO axial-vector isovector and isoscalar coupling constants, respectively, and a summation over repeated $i=u+d$ and $s$ for the index of isoscalar couplings is implied. In order to derive the full axion-nucleon and photon-nucleon couplings up to $\mathcal{O}\left(p^2\right)$, the Lagrangians have to be expressed in terms of the axion field $a$ and the matrix-valued field $u$ has to be expanded to the required order. For the calculation of the $\gamma N\to a N$ amplitudes up to $\mathcal{O}\left(p^2\right)$, we can set $u=\mathds{1}$ and expand the exponentials in $\mathcal{M}_a$ up to $\mathcal{O}(a)$. The relevant chiral building blocks in Eqs.~\eqref{eq.lagan1} and \eqref{eq.lagan2} to the present calculation are 
\begin{align}\label{eq.current}
    \begin{aligned}
    D_\mu & =\partial_\mu-ie Q_N A_\mu\ , \qquad   F^{+}_{\mu\nu}  =2eQ_N\left(\partial_\mu A_\nu-\partial_\nu A_\mu\right)\ ,  \\
    u_\mu & =c_{u-d} \frac{\partial_\mu a}{f_a} \tau^3\ , \qquad 
    \tilde{u}_{\mu,i=(u+d,s)}  =c_{i=(u+d,s)} \frac{\partial_\mu a}{f_a} \mathds{1}\ . 
    \end{aligned}
\end{align} 
In the phenomenological discussions, we will take the physical (isospin-averaged) value to estimate the LO baryon mass $m_N=938.92~$MeV, as the difference between the LO and physical masses only amounts to the effects of higher orders than those considered here. The isospin-averaged pion mass will be also taken as $m_\pi = 138.03$~MeV. Following the method given in Ref.~\cite{GrillidiCortona:2015jxo}, one can  determine the values of $g_A$ and $g_{0}^{(u+d,s)}$ by matching the single-nucleon matrix element of the quark current operators in Eq.~\eqref{eq.L3} and the effective axion-nucleon ones in Eq.~\eqref{eq.lagan1} by using the relation $\langle N |\bar{q}\gamma^\mu\gamma_5q | N\rangle = s^\mu \Delta q$, with $s^\mu$ the spin vector of the nucleon  $N$~\cite{Manohar:2000dt,Barone:2001sp,GrillidiCortona:2015jxo}. This procedure relates the couplings of $g_A$ and $g_{0}^{(u+d,s)}$ with the nucleon matrix elements $\Delta q, (q=u,d,s)$, whose explicit forms read~\cite{GrillidiCortona:2015jxo} 
\begin{align}\label{eq.gag0}
    \begin{aligned}
    g_A & =\Delta u-\Delta d\ , \qquad   g_{0}^{u+d} =\Delta u+\Delta d\ , \qquad  g_{0}^{s} =\Delta s\ .
\end{aligned}
\end{align}
The proton and neutron matrix elements are related by isospin symmetry, i.e. $\left\langle p\left|\bar{u} \gamma^\mu \gamma_5 u\right| p\right\rangle=\left\langle n\left|\bar{d} \gamma^\mu \gamma_5 d\right| n\right\rangle,\left\langle p\left|\bar{d} \gamma^\mu \gamma_5 d\right| p\right\rangle=$ $\left\langle n\left|\bar{u} \gamma^\mu \gamma_5 u\right| n\right\rangle$, and $\left\langle p\left|\bar{s} \gamma^\mu \gamma_5 s\right| p\right\rangle=\left\langle n\left|\bar{s} \gamma^\mu \gamma_5 s\right| n\right\rangle$. The effective couplings of the axion and proton/neutron can be inferred from the combination of $g_{aN}=g_A c_{u-d} \tau^3+g_0^i c_i \mathds{1}$, and the explicit forms are~\cite{GrillidiCortona:2015jxo,Vonk:2020zfh, Vonk:2021sit}
\begin{align}\label{eq.gap}
    g_{ap}&=-\frac{\Delta u+z \Delta d+w \Delta s}{1+z+w}+\Delta u X_u+\Delta d X_d+ \Delta s X_s\ ,\\
    g_{an}&=-\frac{z \Delta u+\Delta d+w \Delta s}{1+z+w}+\Delta d X_u+\Delta u X_d+ \Delta s X_s  \ .\label{eq.gan}
\end{align}
We take the results of the nucleon matrix elements calculated on lattice QCD with $N_f=2+1$ in the isospin limit and the ratios of quark masses from the recent Flavour Lattice Averaging Group (FLAG) review~\cite{FlavourLatticeAveragingGroupFLAG:2021npn} 
\begin{align}
    \Delta u  =0.847(50)\ ,\, \Delta d =-0.407(34)\ ,\, \Delta s=-0.035(13)\ ,\, z=0.485(19)\ ,\, w=0.025(1)\,. 
\end{align}
With such inputs, the model-independent parts of the axion-nucleon couplings in Eqs.~\eqref{eq.gap} and \eqref{eq.gan}, i.e., by taking $X_u=X_d=X_s=0$, are found to be 
\begin{equation}\label{eq.gapgan}
g_{ap}=-0.430(36)\, , \qquad g_{an}=-0.002(30) \, ,
\end{equation}
indicating that the model-independent part of the axion-neutron coupling is much suppressed, comparing with the axion-proton one. However, when the model-dependent parameters $X_{u,d,s}$ are included, such severe suppression generally does not hold anymore.

A remark about the subtle isospin breaking effect is in order. Such effect in our study is introduced in the quark axial transformation matrix $\mathcal{Q}_a$ in Eq.~\eqref{eq.qa}, which in turn contributes to the coefficient of the axial-vector current~\eqref{eq.current1}. It is easy to verify that the coefficient can vary around $15\%$ when turning on/off the strong isospin breaking effect, consistent with the factor $(m_d-m_u)/(m_d+m_u)$ appearing in the axion related amplitudes, e.g. see a recent comprehensive study of axion-hadron interactions in Ref.~\cite{GrillidiCortona:2015jxo}. For the nonperturbative matrix elements of the axial-vector currents, such as those in Eq.~\eqref{eq.gag0} and the isospin relations below this equation, we take the results from the lattice QCD simulations performed in the isospin limit, implying that the isospin breaking effect is ignored for these matrix elements. According to the experience in hadron physics, the isospin breaking typically brings a correction at the level around $1\sim 3\%$~\cite{Amoros:2001cp,Hoferichter:2009ez}, which can be roughly estimated by $\epsilon/\Lambda_\chi^2$, with $\epsilon=B(m_d-m_u)$ and $\Lambda_\chi\sim m_\rho$. Therefore we consider that it is a reasonable approximation to neglect the isospin breaking effects in the hadronic matrix elements.

From the above discussions, we can derive the $\mathcal{O}(p)$ Feynman rules of the $NN\gamma$ and $NNa$ vertices,   
\begin{eqnarray}
\label{eq.vnng1}
    v_\mu(\gamma_\mu N\to N)=ieQ_N\gamma_\mu\,,\\
    v\left(N\to a(q)N\right)=-\frac{g_{aN}}{2f_a}\slashed{q}\gamma_5\,,
\label{eq.vnna}
\end{eqnarray}
and the $\mathcal{O}(p^2)$ Feynman rule of the $NN\gamma$ vertex,
\begin{align}\label{eq.vnng2}
\begin{aligned}
    v_\mu(\gamma_\mu(q)N\to N)=-\frac{e\left(c_6 Q_N+c_7\right)}{2m_N}\sigma_{\mu\nu}q^\nu\,.
\end{aligned}
\end{align}
The LECs $c_6$ and $c_7$ are related to the anomalous magnetic moments of nucleons via $c_6=\kappa_p-\kappa_n, c_7=\kappa_n$, where $\kappa_p$ and $\kappa_n$ are the anomalous magnetic moments of proton and neutron, respectively. According to the PDG results~\cite{ParticleDataGroup:2022pth}, $\kappa_p=1.793,\kappa_n=-1.913$, we can determine 
\begin{equation}
c_6=3.706\,, \qquad c_7=-1.913 \,. 
\end{equation}

\begin{figure}[h]
    \centering
    \includegraphics[width=.6\textwidth]{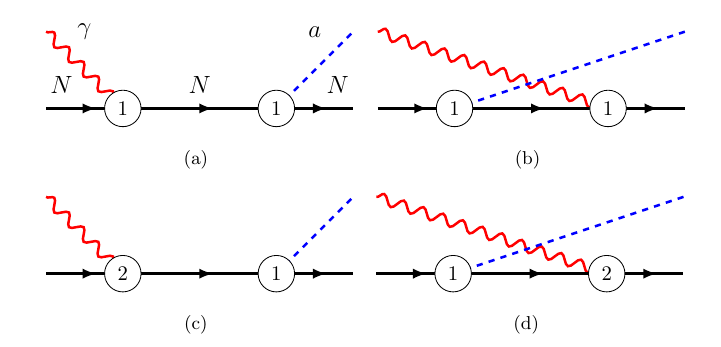} 
    \caption{Tree-level nucleon-mediated diagrams for $\gamma N\to a N$ arise from the $\mathcal{O}(p)$  and $\mathcal{O}(p^2)$ chiral Lagrangians, whose vertices are represented by the circles with number 1 and 2, respectively.}\label{fig.Feyn}
\end{figure}

Next we calculate the $\gamma(k_\gamma) + N(k_N) \to a(p_a) + N(p_N)$ amplitudes up to $O(p^2)$ by using the interaction vertices given in Eqs.~\eqref{eq.vnng1}-\eqref{eq.vnng2}. The relevant Feynman diagrams are shown in Fig.~\ref{fig.Feyn}. 
For the $s$- and $u$-channel nucleon-mediated Feynman diagrams at $O(p)$, i.e., the diagrams (a) and (b) of Fig.~\ref{fig.Feyn}, the expression is 
\begin{align}
    i\mathcal{M}_N^{(1)}=&\epsilon^\mu\bar{N}\left[\left(-\frac{g_{aN}}{2f_a}\slashed{p}_a\gamma_5\right)\frac{i}{\slashed{k}_\gamma+\slashed{k}_N-m_N}ieQ_N\gamma_\mu+ieQ_N\gamma_\mu\frac{i}{\slashed{p}_N-\slashed{k}_\gamma-m_N}\left(-\frac{g_{aN}}{2f_a}\slashed{p}_a\gamma_5\right)\right]N\ ,
\end{align}
and for the Feynman diagrams at $O(p^2)$, i.e., the diagrams (c) and (d) of Fig.~\ref{fig.Feyn}, the corresponding formula is  
\begin{align}
    i\mathcal{M}_N^{(2)}=&\epsilon^\mu\bar{N}\left[\left(-\frac{g_{aN}}{2f_a}\slashed{p}_a\gamma_5\right)\frac{i}{\slashed{k}_\gamma+\slashed{k}_N-m_N}\left(-\frac{e\left(c_6 Q_N+c_7\right)}{2m_N}\sigma_{\mu\nu}k_\gamma^\nu\right) \right.\nonumber\\
    &\left.+\left(-\frac{e\left(c_6 Q_N+c_7\right)}{2m_N}\sigma_{\mu\nu}k_\gamma^\nu\right)\frac{i}{\slashed{p}_N-\slashed{k}_\gamma-m_N}\left(-\frac{g_{aN}}{2f_a}\slashed{p}_a\gamma_5\right)\right]N\,.
\end{align}

\section{Contributions of the anomalous axion-photon interactions }\label{sec.agg}

\begin{figure}[h]
    \centering
    \includegraphics[width=.6\textwidth]{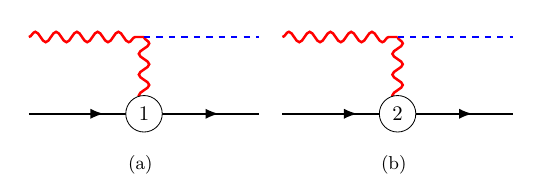}
    \caption{Tree-level  photon-mediated diagrams for $\gamma N\to a N$ arising from the $\mathcal{O}(p)$ and $\mathcal{O}(p^2)$ chiral Lagrangians.}\label{fig.Primakoff}
\end{figure}

In addition to the nucleon-mediated contributions to the $\gamma N\to a N$ amplitudes shown in Fig.~\ref{fig.Feyn}, the anomalous axion-photon-photon vertex also enters the axion photoproduction amplitudes, i.e., the so-called Primakoff production amplitudes, which are shown in Fig.~\ref{fig.Primakoff}. 
The axion-photon-photon interacting operator reads 
\begin{align}\label{eq.lagagg}
    \mathcal{L}_{a\gamma}=\frac{\alpha}{8 \pi} \frac{C_{a \gamma}}{f_a} a F^{\mu\nu} \tilde{F}_{\mu\nu}\ , 
\end{align}
with $F^{\mu\nu}$ and $\tilde{F}_{\mu \nu}=\frac{1}{2} \epsilon_{\mu \nu \alpha \beta} F^{\alpha \beta}$ the photon field strength and its dual, and $\alpha=e^2/(4\pi)$ stands for the fine structure constant. After taking into account the $SU(2)$ higher-order chiral corrections, the updated determination in Ref.~\cite{GrillidiCortona:2015jxo} gives 
\begin{align}\label{eq.gapp}
C_{a\gamma}=\frac{E}{N}-1.92(4) \,,
\end{align}
where $N$ and $E$ are the model-dependent QCD and QED anomaly coefficients, respectively, and the value of $-1.92(4)$ corresponds to the model-independent part of the axion-photon coupling at next-to-leading order (NLO) from the $SU(2)$ chiral EFT. Actually, the value $-2.05(3)$~\cite{Lu:2020rhp} in the $SU(3)$ version is very similar to the one in Eq.~\eqref{eq.gapp}. In later uncertainty discussions, we will take the difference of the central values between the $SU(2)$ and $SU(3)$ chiral EFT as an additional systematical uncertainty, which will be added quadratically with the one of 0.04 in Eq.~\eqref{eq.gapp}. As a result, the model-independent part of $C_{a\gamma}$ will be taken as $-1.92(14)$ for later discussions.

For the $t$-channel photon-mediated diagrams in Fig.~\ref{fig.Primakoff}, the corresponding amplitudes at $\mathcal{O}(p)$ and $\mathcal{O}(p^2)$ take the forms 
\begin{align}
    i\mathcal{M}_\gamma^{(1)}=&i\frac{\alpha C_{a\gamma}}{2\pi f_a}\epsilon_{\alpha\beta\mu\nu}\epsilon^\mu k_\gamma^\alpha p_a^\beta \left(-i\frac{g^{\nu \rho}}{\left(k_\gamma-p_a\right)^2}\right)\Bar{N}ieQ_N\gamma_\rho N\ ,\\
    i\mathcal{M}_\gamma^{(2)}=&i\frac{\alpha C_{a\gamma}}{2\pi f_a}\epsilon_{\alpha\beta\mu\nu}\epsilon^\mu k_\gamma^\alpha p_a^\beta \left(-i\frac{g^{\nu \rho}}{\left(k_\gamma-p_a\right)^2}\right)\Bar{N}\left(-\frac{e\left(c_6 Q_N+c_7\right)}{2m_N}\sigma_{\rho\kappa}\left(k_\gamma-p_a\right)^\kappa\right) N\,,
\end{align}
where the superscripts $(1)$ and $(2)$ denote the chiral orders. 

Furthermore, comparing with the Feynman diagrams governed by the $a\gamma\gamma$ vertex in Fig.~\ref{fig.Primakoff}, one would expect that the $\rho a\gamma$ and $\omega a \gamma$ interacting vertices could also play relevant roles in the $\gamma N\to a N$ processes, due to the obvious enhancements of the $\rho NN$ and $\omega NN$ couplings over the $\gamma NN$ ones, as recently discussed in Ref.~\cite{Chakraborty:2024tyx}. 
Therefore we consider another type of important contribution from the $t$-channel vector-meson exchanges. An alternative related issue is about the potential contribution from the excited baryon states, among which the nearest one is the $\Delta(1232)$ resonance. However, the $aN\Delta$ coupling is suppressed by the isospin breaking factor $m_u-m_d$~\cite{Vonk:2022tho}. Furthermore, the $\gamma N\Delta$ interacting vertex starts at $O(p^2)$~\cite{GuerreroNavarro:2019fqb}. Therefore the effect of the $\Delta$ resonance is expected to be suppressed and we will not explicitly include it in the present study.

\begin{figure}[h]
    \centering
    \includegraphics[width=.4\textwidth]{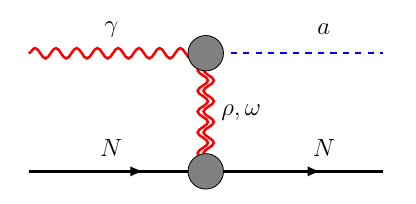}
    \caption{Vector-meson-mediated diagrams for $\gamma N\to a N$.}\label{fig.VM1}
\end{figure}

The relevant lowest-lying vector mesons in the $t$ channel are $\rho$ and $\omega$,~\footnote{According to the Okubo-Zweig-Iizuka rule, the $\phi NN$ coupling is expected to be much suppressed and we do not take the $\phi$ exchange into account in this work.} and the corresponding diagram is shown in Fig.~\ref{fig.VM1}. To proceed with the calculation of such Feynman diagrams, we need to provide two types of interacting vertices: $V a\gamma$ and $VNN$, with $V=\rho,\omega$. It is noted that very recently the $Va\gamma$ couplings have been calculated in the framework of hidden local symmetry in Ref~\cite{Bai:2024lpq}. 

Our strategy to estimate the values of the $Va\gamma$ couplings relies on the use of the effective $VP\gamma$ couplings $g_{\small VP\gamma}$ and the mixing strengths $\theta_{aP}$ between the axion and the light pseudoscalar mesons $P=\pi^0,\eta,\eta'$, i.e.,
\begin{equation}\label{eq.gapcoupling}
g_{\small Va\gamma}=\sum_{P=\pi^0,\eta,\eta'} \theta_{aP} g_{\small VP\gamma}\,.
\end{equation}  
We should mention that the way to construct the $Va\gamma$ vertex is somewhat different from the calculation of the $aNN$ ones in the previous section. Instead of performing the quark axial transformation with $\mathcal{Q}_a$ in Eq.~\eqref{eq.qa} and explicitly writing out the axial currents with vector resonances, we will keep the preexisting mixing terms between the axion and light pseudoscalar mesons ($\pi^0,\eta,\eta'$) whose strengths are given by $\theta_{aP}$ in Eq.~\eqref{eq.gapcoupling}, and assume that the $Va\gamma$ couplings are given by the product of $\theta_{aP}$ and the $VP\gamma$ couplings.

\begin{figure}[h]
    \centering
    \includegraphics[width=.9\textwidth]{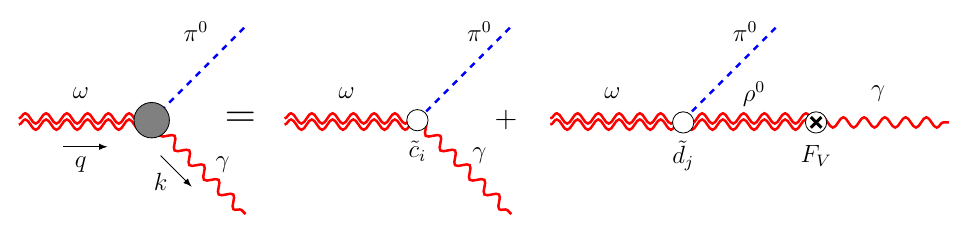}
    \caption{Illustration of the Feynman diagrams for the process $\omega\to \pi^0\gamma$. The definitions of the $VJP$ couplings ($\tilde{c}_i$), the $VVP$ ones ($\tilde{d}_j$) and the $\rho$-$\gamma$ transition strength $F_V$ are given in Refs.~\cite{Ruiz-Femenia:2003jdx,Chen:2012vw}. }\label{fig.Omega_pigamma}
\end{figure}

Regarding the $VP\gamma$ interactions, they have been extensively studied in the framework of resonance chiral theory (R$\chi$T)~\cite{Ruiz-Femenia:2003jdx,Chen:2012vw,Chen:2014yta,Yan:2023bwt,Roig:2013baa,Qin:2020udp,Guo:2010dv,Chen:2022nxm,Arteaga:2022xxy}. Not only many radiative decay processes involving light-flavor mesons~\cite{Ruiz-Femenia:2003jdx,Chen:2012vw,Chen:2014yta,Yan:2023bwt,Roig:2013baa,Qin:2020udp} but also the relevant $\tau$ decay channels~\cite{Guo:2010dv,Chen:2022nxm,Arteaga:2022xxy} have been successfully described by the $VJP$ (being $J$ the external source field) and $VVP$ operators from R$\chi$T. We take the $\omega\to\pi^0\gamma$ process as an example to illustrate the R$\chi$T calculation in Fig.~\ref{fig.Omega_pigamma}. Comparing with the phenomenological vector-meson-dominance (VMD) model, general interacting operators that respect global symmetries of QCD are constructed in the R$\chi$T study, including both the direct $VP\gamma$ vertices (denoted by $\tilde{c}_i$) and the vector-meson-mediated ones (denoted by $\tilde{d}_j$). For the constructions of R$\chi$T Lagrangians, we refer to Refs.~\cite{Ruiz-Femenia:2003jdx,Chen:2012vw} for further details.

Here we will directly take the results from the R$\chi$T study to describe the on-shell $VP\gamma$ interacting vertices~\cite{Ruiz-Femenia:2003jdx,Chen:2012vw,Chen:2014yta,Yan:2023bwt}
\begin{align}
i\mathcal{M}_{VP \gamma}= ie\epsilon_{\alpha \beta \rho \sigma} \epsilon_V^\alpha(q) \epsilon_\gamma^{*\beta}(k) q^\rho k^\sigma \mathcal{F}_{VP\gamma^*}(Q^2=0)\ ,
\end{align}
where $q$ and $k$ denote the four-momenta of the vector meson and photon in order, the transferred momentum squared is $Q^2=-k^2$, and $\epsilon_V(q)$ and $\epsilon_\gamma(k)$ correspond to the polarization vectors of the vector meson and photon, respectively. $\mathcal{F}_{VP\gamma^*}(Q^2=-k^2)$ stands for the general form factor when involving the off-shell photon, and for the on-shell real photon one can take $Q^2=0$. The effective $VP\gamma$ couplings $g_{\small VP\gamma}$ in Eq.~\eqref{eq.gapcoupling} are simply given by the on-shell form factor, i.e., $g_{\small VP\gamma}=\mathcal{F}_{VP\gamma^*}(Q^2=0)$. The latter quantities have been calculated for various processes with $V=\rho,\omega,\phi,K^*$ and $P=\pi,K,\eta,\eta'$ in Ref.~\cite{Chen:2012vw}. We explicitly give pertinent expressions to the current study for the sake of completeness,   
\begin{align}\label{eq.Fomega1}
    g_{\omega \pi^0\gamma}\equiv\mathcal{F}_{\omega\pi^0\gamma^*}(0)= & \frac{2 \sqrt{2}}{F_\pi M_V m_\omega}\left[\left(\tilde{c}_1+\tilde{c}_2+8 \tilde{c}_3-\tilde{c}_5\right) m_\pi^2+\left(\tilde{c}_2+\tilde{c}_5-\tilde{c}_1-2 \tilde{c}_6\right) m_\omega^2\right] \nonumber\\
    & -\frac{4 F_V}{F_\pi m_\rho^2 m_\omega}\left[\left(\tilde{d}_1+8 \tilde{d}_2-\tilde{d}_3\right) m_\pi^2+\tilde{d}_3 m_\omega^2\right]\ ,
\end{align}
\begin{align}
    g_{\rho^0\pi^0\gamma}\equiv\mathcal{F}_{\rho^0\pi^0 \gamma^*}(0)  = & \frac{2 \sqrt{2}}{3F_\pi M_V m_\rho}\left[\left(\tilde{c}_1+\tilde{c}_2+8 \tilde{c}_3-\tilde{c}_5\right) m_\pi^2+\left(\tilde{c}_2+\tilde{c}_5-\tilde{c}_1-2 \tilde{c}_6\right) m_\rho^2\right] \nonumber\\
    & -\frac{4 F_V}{3F_\pi m_\omega^2 m_\rho}\left[\left(\tilde{d}_1+8 \tilde{d}_2-\tilde{d}_3\right) m_\pi^2+\tilde{d}_3 m_\rho^2\right] \ ,
\end{align}
\begin{align}
    g_{\rho^0 \eta\gamma}\equiv\mathcal{F}_{\rho \eta \gamma^*}(0)= & \frac{4}{M_V m_\rho F_\pi} a_1\left[m_\rho^2\left(\tilde{c}_2-\tilde{c}_1+\tilde{c}_5-2 \tilde{c}_6\right)+m_\eta^2\left(\tilde{c}_2+\tilde{c}_1-\tilde{c}_5\right)+8 \tilde{c}_3 m_\pi^2\right] \nonumber\\
    & -\frac{4 \sqrt{2} F_V}{m_\rho^3 F_\pi} a_1\left[\tilde{d}_3\left(m_\rho^2-m_\eta^2\right)+\tilde{d}_1 m_\eta^2+8 \tilde{d}_2 m_\pi^2\right] \nonumber\\
    & +\frac{\sin \theta_8}{\cos \left(\theta_0-\theta_8\right) F_0}\left(-\frac{4 \sqrt{2} M_V}{m_\rho} \tilde{c}_8+\frac{8 F_V M_V^2}{m_\rho^3} \tilde{d}_5\right)\ ,
\end{align}
\begin{align}    
    g_{\rho^0 \eta^\prime\gamma}\equiv F_{\rho \eta^{\prime} \gamma^*}(0)= & \frac{4}{M_V m_\rho F_\pi} a_2\left[m_\rho^2\left(\tilde{c}_2-\tilde{c}_1+\tilde{c}_5-2 \tilde{c}_6\right)+m_{\eta^{\prime}}^2\left(\tilde{c}_2+\tilde{c}_1-\tilde{c}_5\right)+8 \tilde{c}_3 m_\pi^2\right] \nonumber\\
    & -\frac{4 \sqrt{2} F_V}{m_\rho^3 F_\pi} a_2\left[\tilde{d}_3\left(m_\rho^2-m_{\eta^{\prime}}^2\right)+\tilde{d}_1 m_{\eta^{\prime}}^2+8 \tilde{d}_2 m_\pi^2\right] \nonumber\\
    & +\frac{\cos \theta_8}{\cos \left(\theta_0-\theta_8\right) F_0}\left(\frac{4 \sqrt{2} M_V}{m_\rho} \tilde{c}_8-\frac{8 F_V M_V^2}{m_\rho^3} \tilde{d}_5\right)\ ,
\end{align}
\begin{align}
    g_{\omega \eta\gamma}\equiv \mathcal{F}_{\omega \eta \gamma^*}(0)= & \frac{4}{3 M_V m_\omega F_\pi} a_1\left[m_\omega^2\left(\tilde{c}_2-\tilde{c}_1+\tilde{c}_5-2 \tilde{c}_6\right)+m_\eta^2\left(\tilde{c}_2+\tilde{c}_1-\tilde{c}_5\right)+8 \tilde{c}_3 m_\pi^2\right] \nonumber\\
    & -\frac{4 \sqrt{2} F_V}{3 m_\omega^3 F_\pi} a_1\left[\tilde{d}_3\left(m_\omega^2-m_\eta^2\right)+\tilde{d}_1 m_\eta^2+8 \tilde{d}_2 m_\pi^2\right] \nonumber\\
    & +\frac{\sin \theta_8}{3 \cos \left(\theta_0-\theta_8\right) F_0}\left(-\frac{4 \sqrt{2} M_V}{m_\omega} \tilde{c}_8+\frac{8 F_V M_V^2}{m_\omega^3} \tilde{d}_5\right)\ , 
\end{align}
\begin{align}\label{eq.Fomegaepg}
    g_{\omega \eta^\prime\gamma}\equiv \mathcal{F}_{\omega \eta^{\prime} \gamma^*}(0)= & \frac{4}{3 M_V m_\omega F_\pi} a_2\left[m_\omega^2\left(\tilde{c}_2-\tilde{c}_1+\tilde{c}_5-2 \tilde{c}_6\right)+m_{\eta^{\prime}}^2\left(\tilde{c}_2+\tilde{c}_1-\tilde{c}_5\right)+8 \tilde{c}_3 m_\pi^2\right] \nonumber\\
    & -\frac{4 \sqrt{2} F_V}{3 m_\omega^3 F_\pi}  a_2\left[\tilde{d}_3\left(m_\omega^2-m_{\eta^{\prime}}^2\right)+\tilde{d}_1 m_{\eta^{\prime}}^2+8 \tilde{d}_2 m_\pi^2\right] \nonumber\\
    & +\frac{\cos \theta_8}{3 \cos \left(\theta_0-\theta_8\right) F_0}\left(\frac{4 \sqrt{2} M_V}{m_\omega} \tilde{c}_8-\frac{8 F_V M_V^2}{m_\omega^3} \tilde{d}_5\right)\ ,
\end{align}
where the abbreviated factors $a_{i=1,2,3,4}$ from the $\eta$-$\eta'$ mixing are defined as  
\begin{align}
    & a_1=\frac{F_\pi}{\cos \left(\theta_0-\theta_8\right)}\left(\frac{1}{\sqrt{6}} \frac{\cos \theta_0}{F_8}-\frac{1}{\sqrt{3}} \frac{\sin \theta_8}{F_0}\right)\ , \\
    & a_2=\frac{F_\pi}{\cos \left(\theta_0-\theta_8\right)}\left(\frac{1}{\sqrt{6}} \frac{\sin \theta_0}{F_8}+\frac{1}{\sqrt{3}} \frac{\cos \theta_8}{F_0}\right)\ , \\
    & a_3=\frac{F_\pi}{\cos \left(\theta_0-\theta_8\right)}\left(-\frac{2}{\sqrt{6}} \frac{\cos \theta_0}{F_8}-\frac{1}{\sqrt{3}} \frac{\sin \theta_8}{F_0}\right)\ , \\
    & a_4=\frac{F_\pi}{\cos \left(\theta_0-\theta_8\right)}\left(-\frac{2}{\sqrt{6}} \frac{\sin \theta_0}{F_8}+\frac{1}{\sqrt{3}} \frac{\cos \theta_8}{F_0}\right)\ ,
\end{align}
with $F_0, F_8,\theta_0,\theta_8$ the parameters in the two-mixing-angle formalism of the $\eta$-$\eta'$ system~\cite{Chen:2012vw}. 
In the R$\chi$T framework, it is customary to exploit the short-distance or high-energy behaviors of the various form factors and Green functions as dictated by QCD, to constrain the hadronic couplings~\cite{Ecker:1989yg,Cirigliano:2006hb,Roig:2013baa}. Many of the couplings $\tilde{c}_i$ and $\tilde{d}_j$ in Eqs.~\eqref{eq.Fomega1}-\eqref{eq.Fomegaepg} have been determined in this way~\cite{Ruiz-Femenia:2003jdx,Chen:2012vw}. The relevant ones to our study read
\begin{align}\label{eq.cidj}
    \begin{aligned}
    & \tilde{c}_1+4 \tilde{c}_3=0\ , \quad \tilde{c}_1-\tilde{c}_2+\tilde{c}_5=0\ , \quad \tilde{c}_5-\tilde{c}_6=\frac{N_C M_V}{64 \sqrt{2} \pi^2 F_V}\ , \\
    & \tilde{c}_8=-\frac{\sqrt{2} M_0^2}{\sqrt{3} M_V^2} \tilde{c}_1\ ,\quad \tilde{d}_1+8 \tilde{d}_2-\tilde{d}_3=\frac{F_\pi^2}{8 F_V^2}\ , \quad \tilde{d}_3=-\frac{N_C M_V^2}{64 \pi^2 F_V^2}\ , 
\end{aligned}
\end{align}
where $N_C=3$ and the mass of the singlet $\eta_0$ is taken as $M_0=0.9~$GeV~\cite{Chen:2012vw}. The vector-meson mass $M_V$ in the chiral limit can be well estimated by the mass of $\rho(770)$.

For the remaining unknown parameters, we will use the fitted values from Refs.~\cite{Yan:2023nqz,Chen:2012vw}. A strong linear correlation between $\tilde{d}_2$ and $\tilde{d}_5$: $\tilde{d}_5=4.4 \tilde{d}_2-0.06$, is observed in the former reference, and we will take this linear correlation in our study as well. The explicit values of the relevant parameters from Refs.~\cite{Yan:2023nqz,Chen:2012vw} are given in Table.~\ref{tab.para1}. In later discussions, the results by taking the values of parameters of Refs.~\cite{Yan:2023nqz} and ~\cite{Chen:2012vw} will be referred as Set-I and Set-II, respectively. 
\begin{table}[tb]
    \centering
    \begin{tabular}{ccc}
    \hline\hline 
    Parameters \,\,\,\,\, & Set-I (Ref.~\cite{Yan:2023nqz})\,\,\,\,\,\,\,\,\,\, & Set-II (Ref.~\cite{Chen:2012vw}) \\
    \hline
$F_8$ & $(1.41 \pm 0.02) F_\pi \, $ & $(1.37 \pm 0.07) F_\pi$ \\
$F_0$ & $(1.36 \pm 0.03) F_\pi$ & $(1.19 \pm 0.18) F_\pi$ \\
$\theta_8$ & $(-24.3\pm0.4)^{\circ}$ & $(-21.1 \pm 6.0)^{\circ}$ \\
$\theta_0$ & $(-12.8\pm0.5)^{\circ}$ & $(-2.5 \pm 8.2)^{\circ}$ \\
$F_V$(GeV) & $0.1390 \pm 0.0002$ & $0.1366 \pm 0.0035$ \\
$\tilde{c}_3$ & $0.0046 \pm 0.0003$ & $0.0109 \pm 0.0161$ \\
$\tilde{d}_2$ & $0.100 \pm 0.008$ & $0.086 \pm 0.085$ \\
    \hline\hline
    \end{tabular}
    \caption{The two sets of values for the relevant parameters from Refs.~\cite{Yan:2023nqz,Chen:2012vw}. For comparison, we provide the results of Ref.~\cite{Chen:2012vw} which are obtained by fitting only to the light hadron radiative decay processes, and the results of Ref.~\cite{Yan:2023nqz} which are obtained by fitting to the light hadrons, $J/\psi$ and $\psi'$ radiative decay processes simultaneously.  }
    \label{tab.para1}
\end{table}

For the mixing parameters $\theta_{aP}$ between the axion and light-flavor mesons in Eq.~\eqref{eq.gapcoupling}, we adopt the mixing pattern in Ref.~\cite{Alves:2024dpa} and after translating into our notation the explicit formulas with vanishing bare axion mass $m_{a,0}=0$ read  
\begin{align}
    \theta_{a\pi}^{(\mathrm{0})} & =-\frac{F_\pi}{f_a}\left[\frac{z X_u-X_d}{1+z}+\frac{1-z}{2(1+z)} \right]+\mathcal{O}(w)\ ,\label{eq.mix1}\\
     \theta_{a \eta_8}^{(\mathrm{0})}   &= \frac{F_\pi}{f_a} \frac{\sqrt{3}}{2}\left(X_s-\frac{1}{3}\right)+\mathcal{O}(w)\ , \qquad    \theta_{a \eta_0}^{(\mathrm{0})}  = -\frac{F_\pi}{\sqrt{6}f_a}+\mathcal{O}(w)   \ . \label{eq.mix2}
\end{align} 
For the situation with a finite bare mass $m_{a,0}$, the mass squared of the physical axion is $m_a^2=m_{a,0}^2+\mathcal{O}(\frac{1}{f_a^2})$. By neglecting the $\pi$-$\eta$ and $\pi$-$\eta^{\prime}$ mixing terms, which are indeed demonstrated to be very small~\cite{Gao:2022xqz}, the axion-meson mixing strengths can be written as~\cite{Alves:2024dpa} 
\begin{align}
    \theta_{a \pi} = & \theta_{a \pi}^{(\mathrm{0})}\left(1+\frac{m_a^2}{m_\pi^2-m_a^2}\right)\ , \\
    \theta_{a \eta_8} = & \theta_{a \eta_8}^{(\mathrm{0})}\left(1+\cos ^2 \theta_{\eta \eta^{\prime}} \frac{m_a^2}{m_\eta^2-m_a^2}+\sin ^2 \theta_{\eta \eta^{\prime}} \frac{m_a^2}{m_{\eta^{\prime}}^2-m_a^2}\right) \nonumber\\
    & +\theta_{a \eta_0}^{(\mathrm{0})} \frac{\sin 2 \theta_{\eta \eta^{\prime}}}{2}\left(\frac{m_a^2}{m_{\eta^{\prime}}^2-m_a^2}-\frac{m_a^2}{m_\eta^2-m_a^2}\right)\ , \\
    \theta_{a \eta_0} = & \theta_{a \eta_0}^{(\mathrm{0})}\left(1+\sin ^2 \theta_{\eta \eta^{\prime}} \frac{m_a^2}{m_\eta^2-m_a^2}+\cos ^2 \theta_{\eta \eta^{\prime}} \frac{m_a^2}{m_{\eta^{\prime}}^2-m_a^2}\right) \nonumber\\
    & +\theta_{a \eta_8}^{(\mathrm{0})} \frac{\sin 2 \theta_{\eta \eta^{\prime}}}{2}\left(\frac{m_a^2}{m_{\eta^{\prime}}^2-m_a^2}-\frac{m_a^2}{m_\eta^2-m_a^2}\right)\ , 
\end{align}
where $\theta_{\eta\eta'}$ denotes the LO $\eta$-$\eta'$ mixing angle~\cite{Guo:2015xva}. 
The mixing strengths of the axion and the $\eta, \eta'$ mesons are then given by 
\begin{equation} \label{eq.mix3}
\theta_{a \eta} \equiv \theta_{a \eta_8}\cos\theta_{\eta\eta'}-\theta_{a \eta_0} \sin\theta_{\eta\eta'} \ , \qquad \theta_{a \eta^\prime}  \equiv \theta_{a \eta_8} \sin\theta_{\eta\eta'}+\theta_{a \eta_0} \cos\theta_{\eta\eta'} \ .
\end{equation}
It is noted that a systematical calculation of the mixing pattern among the $\pi^0,\eta,\eta'$ and axion has been carried out up to NLO in $U(3)$ chiral EFT by only including the model-independent axion interactions, namely by taking $X_{u}=X_d=X_s=0$, in Ref.~\cite{Gao:2022xqz}. Since we consider the model-dependent axion-quark couplings in this work, we will use the neat mixing pattern given in Eqs.~\eqref{eq.mix1}-\eqref{eq.mix3}. 
For the LO $\eta$-$\eta^\prime$ mixing angle, we will take its value of $\theta_{\eta \eta^{\prime}}=-19.6^{\circ}$ from the recent determination in Ref.~\cite{Gao:2022xqz}.

The $Va\gamma$ interacting vertices can be written as 
\begin{align}
i\mathcal{M}_{Va \gamma}= ie g_{Va\gamma} \epsilon_{\alpha \beta \rho \sigma} \epsilon_V^\alpha(q) \epsilon_\gamma^{*\beta}(k) q^\rho k^\sigma \ ,
\end{align}
where the $g_{Va\gamma}$ couplings can be calculated via Eq.~\eqref{eq.gapcoupling}. By taking $X_u=X_d=X_s=0, m_a=0$ and the parameters in Table.~\ref{tab.para1}, one can obtain the model-independent couplings for $g_{\rho^0 a\gamma}, g_{\omega a\gamma}$, whose explicit values are
\begin{align}\label{eq.vag2num}
    f_a g_{\rho^0 a\gamma}=-0.098(2)\ ,\quad f_a g_{\omega a\gamma}=-0.066(2)\ , 
\end{align}
for Set-I with the parameters from Ref.~\cite{Yan:2023nqz}, and 
\begin{align}\label{eq.vag1num}
    f_a g_{\rho^0 a\gamma}=-0.132(20)\ ,\quad f_a g_{\omega a\gamma}=-0.077(7)\ , 
\end{align}
for Set-II with the parameters from Ref.~\cite{Chen:2012vw}.

To proceed with the calculation of the vector-meson-mediated Feynman diagrams in Fig.~\ref{fig.VM1}, one should provide the vector-nucleon-nucleon ($VNN$) interacting vertices as well. We will take the phenomenological descriptions of such vertices in Refs.~\cite{Bauer:2012pv,Janssen:1996kx,Ronchen:2012eg}, which read
\begin{align}\label{eq.L_rhoNN}
    \mathcal{L}_{VNN}=\Bar{N}\bigg( g_{\rho NN}\vec{\rho}_\mu\cdot\vec{\tau} +g_{\omega NN}\omega_\mu \bigg)\gamma^\mu N+\frac{G_\rho}{2}\Bar{N}\vec{\rho}_{\mu\nu}\cdot \vec{\tau}\sigma^{\mu\nu}N\ ,
\end{align}
with $\vec{\rho}_{\mu \nu}=\partial_\mu \vec{\rho}_\nu-\partial_\nu \vec{\rho}_\mu$. We can then calculate the amplitudes corresponding to the Feynman diagrams in Fig.~\ref{fig.VM1}, which are given by
\begin{align}
    i\mathcal{M}_V= &-ie\epsilon_{\mu\nu\alpha\beta}k^\mu_\gamma p^\nu_a\epsilon_\gamma^\alpha \Bar{N} \left(\frac{g_{\rho NN} g_{\rho^0 a\gamma}\tau^3}{m_\rho^2-(k_\gamma-p_a)^2}+\frac{g_{\omega NN} g_{\omega a\gamma}}{m_\omega^2-(k_\gamma-p_a)^2}\right) \gamma^\beta N  \nonumber \\ &
    + eg_{\rho^0 a\gamma} G_\rho \epsilon_{\mu\nu\alpha\beta}k^\mu_\gamma p^\nu_a \epsilon^\alpha_\gamma\Bar{N}\frac{\sigma^{\beta\delta}(k_\gamma-p_a)_\delta\tau^3}{m_\rho^2-(k_\gamma-p_a)^2} N \ .
\end{align} 
Regarding the value of the coupling $g_{\rho NN}$, one could rely on the universality assumption of the $\rho$ couplings and the original Kawarabayashi-Suzuki-Riazuddin-Fayyazuddin (KSRF) relation~\cite{Kawarabayashi:1966kd,Riazuddin:1966sw} to estimate $g_{\rho NN}=g_\rho = \frac{m_\rho}{2\sqrt{2}F_\pi}$. For example, by taking the physical values for the $\rho$ mass $m_\rho=775.3$~MeV and the pion decay constant $F_\pi=92.1$~MeV~\cite{ParticleDataGroup:2022pth}, one obtains $g_{\rho NN} = \frac{m_\rho}{\sqrt{2}F_\pi}\simeq 2.98$~\cite{Bauer:2012pv}. However, if one uses the pion decay constant in the chiral limit $F\simeq 82$~MeV~\cite{Gu:2018swy} to replace its physical value, the former relation gives $g_{\rho NN}\simeq 3.34$,  which is close to the result $g_{\rho NN}=3.25$ obtained in the J\" ulich model~\cite{Janssen:1996kx,Ronchen:2012eg}. The definition of this coupling in Ref.~\cite{Bauer:2012pv} differs from the one in Refs.~\cite{Janssen:1996kx,Ronchen:2012eg} by a factor of 2 that we have properly taken into account here.  For the tensor coupling $G_\rho$ in Eq.~\eqref{eq.L_rhoNN}, the J\" ulich model determines $G_\rho=-10.6~\mathrm{GeV}^{-1}$~\cite{Janssen:1996kx}, which will be used in our later study. The coupling $g_{\omega NN}=11.7$ obtained in the J\" ulich model~\cite{Ronchen:2012eg} is utilized; while for the tensor coupling of the $\omega$, it is usually set to zero in phenomenological discussions~\cite{Janssen:1996kx,Ronchen:2012eg}.

\section{Phenomenological discussions}\label{sec.phenon}

Before stepping into the details of phenomenological discussions, we first set up the notations of the kinematical variables for the $\gamma(k_\gamma)+N(k_N)\to a(p_a)+N(p_N)$ processes, where $N$ stands for either proton or neutron. As usual, we define the usual Mandelstam variables:
\begin{align}
    s=W^2=(k_\gamma+k_N)^2,\quad t=(k_\gamma-p_a)^2,\quad u=(k_\gamma-p_N)^2\ ,
\end{align}
which fulfill the on-shell relation $s+t+u=2 m_N^2+m_a^2$. In later discussions, we use the center of mass (c.m.) frame, in which the three-momenta obey the relation $\mathbf{k}_\gamma+\mathbf{k}_N=\mathbf{p}_a+\mathbf{p}_N=0$. In this frame, the relations of the three-momenta and energies of the involved particles with the c.m. energy squared $s$ read 
\begin{align}
    &|\mathbf{k}_\gamma|=|\mathbf{k}_N|\equiv |\mathbf{k}|=\frac{\sqrt{\lambda\left(s, m_N^2, m_\gamma^2\right)}}{2 \sqrt{s}}\ ,\quad \left|\mathbf{p}_a\right|=\left|\mathbf{p}_N\right|\equiv |\mathbf{p}|=\frac{\sqrt{\lambda\left(s, m_N^2, m_a^2\right)}}{2 \sqrt{s}}\ ,\\
    &E_{\mathbf{k}_N}=\frac{s+m_N^2-m_\gamma^2}{2 \sqrt{s}}\ , \quad E_{\mathbf{p}_N}=\frac{s+m_N^2-m_a^2}{2 \sqrt{s}}\ ,\\
    &E_{\mathbf{k}_\gamma}=\frac{s-m_N^2+m_\gamma^2}{2 \sqrt{s}}\ , \quad E_{\mathbf{p}_a}=\frac{s-m_N^2+m_a^2}{2 \sqrt{s}} 
    \ ,
\end{align}
with the K\"all\'en function $\lambda(a, b, c)=a^2+b^2+c^2-2 a b-2 a c-2 b c$.  

With such preparations, the two-body total cross section can be written as
\begin{align}
    \sigma_{\gamma N\to a N}(s)=\int_{t_1}^{t_0} \md t~\frac{1}{64 \pi s} \frac{1}{\left|\mathbf{k}\right|^2}|\mathcal{\overline{M}}|^2(s,t)\ ,
\end{align}
where $|\mathcal{\overline{M}}|^2(s,t)$ corresponds to the amplitude squared after the spin average and sum of the initial and final states, respectively, and the kinematical boundaries of the $t$ variable are   
\begin{align}
    t_0\left(t_1\right)=\frac{m_a^4}{4 s}-\left(\left|\mathbf{k}\right| \mp \left|\mathbf{p}\right|\right)^2\ .
\end{align}

In order to reveal the relative strengths of different mechanisms in the $\gamma N \to a N$ processes, we distinguish the following different scenarios in later discussions: 
\begin{align}\label{eq.csfour}
    \textbf{I}:\quad &|\mathcal{\overline{M}}|^2=|\overline{\mathcal{M}_N^{(1+2)}}|^2\ , \nonumber\\
    \textbf{II}:\quad &|\mathcal{\overline{M}}|^2=|\overline{\mathcal{M}_\gamma^{(1+2)}}|^2\ ,\nonumber\\
    \textbf{III}:\quad &|\mathcal{\overline{M}}|^2=|\overline{\mathcal{M}_V}|^2\ ,\nonumber\\
    \textbf{IV}:\quad &|\mathcal{\overline{M}}|^2=|\overline{\mathcal{M}_N^{(1+2)}+\mathcal{M}_\gamma^{(1+2)}+\mathcal{M}_V}|^2\ ,
\end{align}
which correspond to the exchanges of nucleon (Fig.~\ref{fig.Feyn}), photon (Fig.~\ref{fig.Primakoff}), vector resonances (Fig.~\ref{fig.VM1}) and their total sums, in order. For each scenario, we give predictions to both the KSVZ and DFSZ axion models by taking several different values for $\tan\beta$ and $E/N$. 

\begin{figure}[tb]
\centering
\includegraphics[width=1\textwidth]{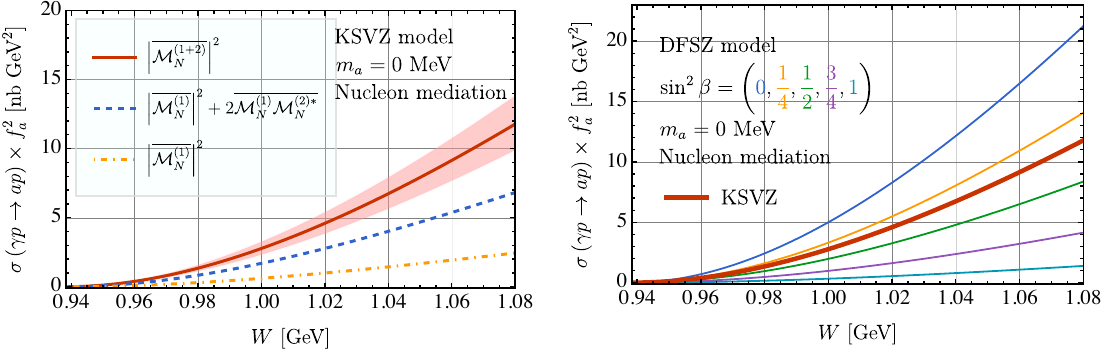} 
\includegraphics[width=1\textwidth]{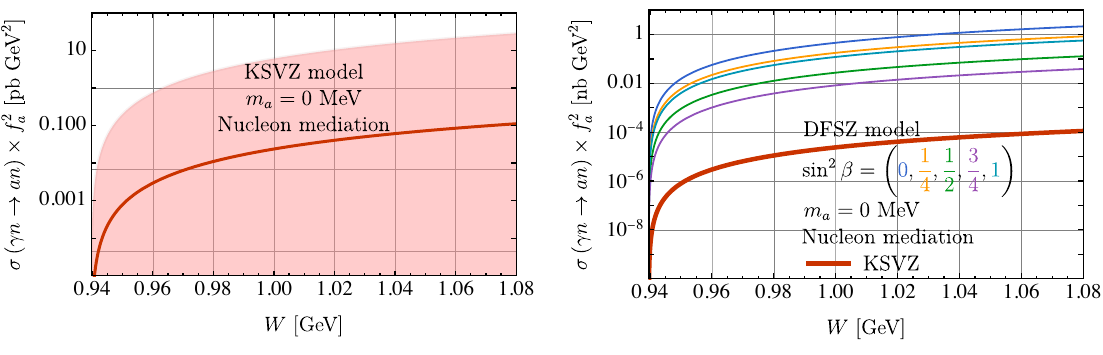} 
\caption{Total cross sections in Scenario {\bf I} for the $\gamma p \to a p$ (top row) and $\gamma n \to a n$ (bottom row)  processes purely contributed by the nucleon-exchange diagrams. Predictions are given up to the c.m. energy $W$ below the $\pi N$ threshold for the KSVZ (left) and DFSZ (right) cases at different values of $\sin ^2 \beta$. In the right panels, the red, blue, orange, green, and purple lines correspond to the DFSZ model parameter $\sin ^2 \beta = 0, 1/4, 1/2, 3/4$ and 1, respectively. In all the panels, $m_a=0~$MeV is utilized. The shaded zones with red color correspond to our estimation of theoretical uncertainties, which are explained in the text. }\label{fig.TCS_1}
\end{figure}

\begin{figure}[tb]
\centering
\includegraphics[width=1\textwidth]{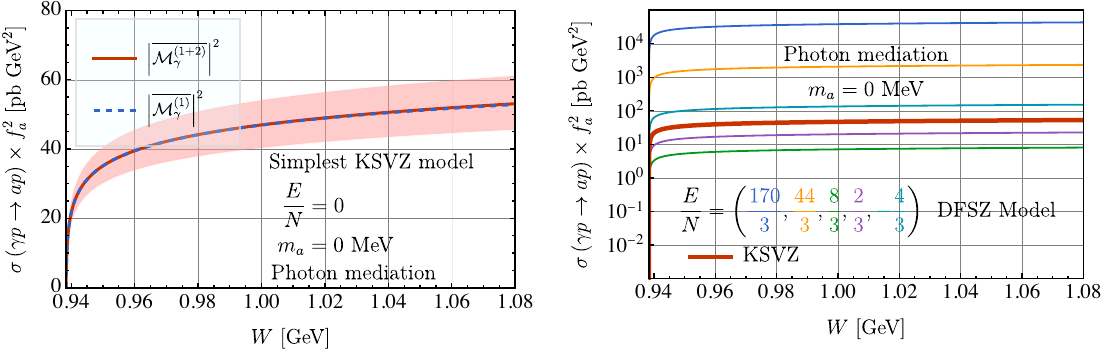} 
\includegraphics[width=1\textwidth]{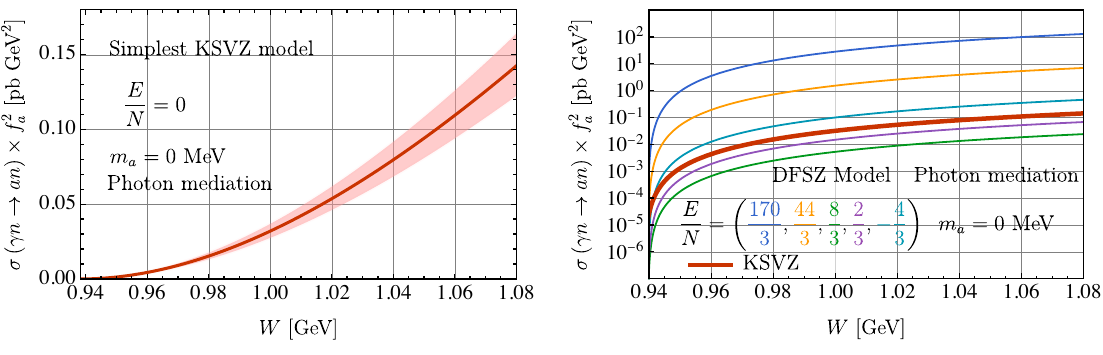} 
\caption{Total cross sections of the photon-mediated $\gamma N \rightarrow a N$ (Primakoff production) as a function of the c.m. energy $W$ below the $\pi N$ threshold for the simplest KSVZ axion (left) and the hadronic axion under a phenomenologically preferred axion window at different valus of $E/N$ (right). Right: the solid lines correspond to the cross sections, in addition, the red, blue, orange, green, and purple lines correspond to the typical parameter $E/N = 170/3, 44/3, 8/3, 2/3$ and $-4/3$~\cite{DiLuzio:2016sbl, DiLuzio:2017pfr}, respectively. Here, we have set $m_a=0~$MeV. The shaded red zones correspond to the uncertainties, see the text for details.}\label{fig.TCS_2}
\end{figure}

\begin{figure}[tb]
\centering
\includegraphics[width=1\textwidth]{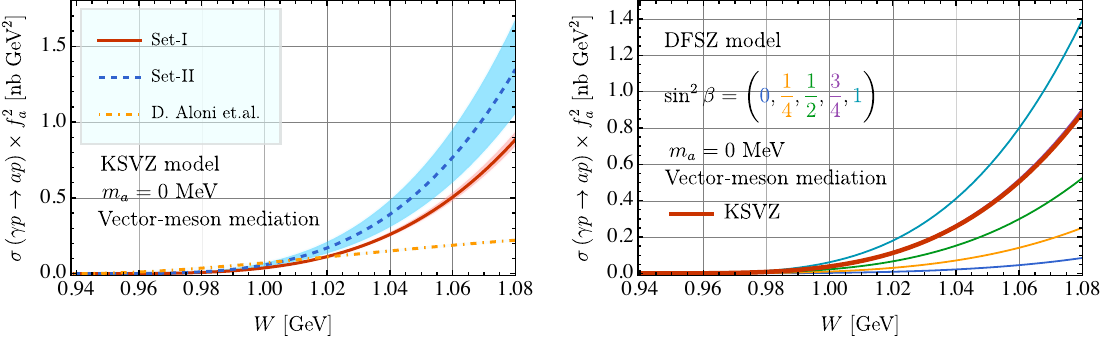} 
\includegraphics[width=1\textwidth]{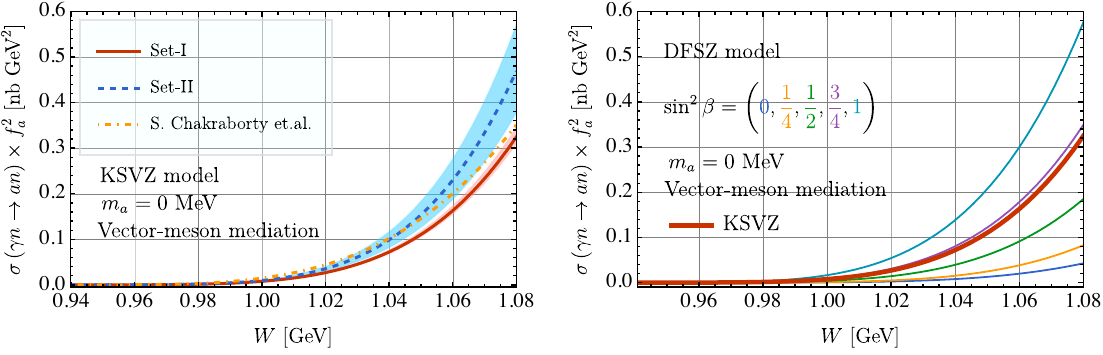} 
\caption{As Fig.~\ref{fig.TCS_1} but for scenarios III. The red solid and blue dashed lines correspond to the parameters used in Refs.~\cite{Yan:2023nqz} and~\cite{Chen:2012vw}, respectively. We also include a dot-dashed yellow line describing the vector-meson-mediated result derived from the VMD model in Refs.~\cite{Aloni:2019ruo,Chakraborty:2024tyx}. The shaded zones correspond to the uncertainties; see the text for details. }\label{fig.TCS_VM1}
\end{figure}

In Figs.~\ref{fig.TCS_1}, \ref{fig.TCS_2} and \ref{fig.TCS_VM1}, we give the cross sections of the $\gamma p \to a p$ and $\gamma n \to a n$ processes by taking $m_a=0$ for Scenarios {\bf I}, {\bf II} and {\bf III}, respectively. 
In Fig.~\ref{fig.TCS_1}, we predict the cross sections to the $\gamma p \to a p$ and $\gamma n \to a n$ processes by setting $m_a=0$ for Scenario {\bf I} that only incorporates the contributions from nucleon exchanges.  The first lesson is that for most of the parameter spaces the cross sections of the neutron channels are much smaller than those of the proton channels. This is due to the fact that $|g_{an}|$ is generally smaller than $|g_{ap}|$, see the discussions around Eqs.~\eqref{eq.gap} and \eqref{eq.gan}. 
For the KSVZ case when only the model-independent axion interaction is included, the NLO correction from the anomalous magnetic moments, i.e. the $c_6$ and $c_7$ terms, to the proton channel is significant, according to the obvious deviation of the red solid line from the dashed yellow line in the top left panel of Fig.~\ref{fig.TCS_1}. In the QCD axion case with $m_a=0$, the expressions of $|\overline{\mathcal{M}_N^{(1+2)}}|^2$ turn out to be very simple, and for the proton channel the explicit forms read 
\begin{align}\label{eq.mn1}
    |\overline{\mathcal{M}_N^{(1)}}|^2 & =-\frac{e^2 g_{ap}^2m_N^2 t^2}{f_a^2(s-m_N^2)(u-m_N^2)}\ ,\\ \label{eq.mn2}
    2\overline{\mathcal{M}_N^{(1)}\mathcal{M}_N^{(2)*}} & =-(c_6+c_7)\frac{e^2 g_{ap}^2m_N^2 t^2}{f_a^2(s-m_N^2)(u-m_N^2)}\ ,\\\label{eq.mn3}
    |\overline{\mathcal{M}_N^{(2)}}|^2 & =-(c_6+c_7)^2\frac{e^2 g_{ap}^2m_N^2 \left(s^2 u^2+m_N^2(s^2+u^2)-4m_N^6(s+u)+5m_N^2\right)}{4f_a^2(s-m_N^2)(u-m_N^2)}\ ,
\end{align}
while for the neutron channel the only nonvanishing term is $|\overline{\mathcal{M}_N^{(2)}}|^2$. 
It is noted that for the proton channel $2\overline{\mathcal{M}_N^{(1)}\mathcal{M}_N^{(2)*}}=(c_6+c_7)|\overline{\mathcal{M}_N^{(1)}}|^2=1.793|\overline{\mathcal{M}_N^{(1)}}|^2$, indicating that the NLO term $2\overline{\mathcal{M}_N^{(1)}\mathcal{M}_N^{(2)*}}$ is comparable to the LO term $|\overline{\mathcal{M}_N^{(1)}}|^2$. It is reminded that the conventional chiral power counting typically refers to the amplitude itself, rather than the amplitude squared. When applying the power counting to the latter quantity, the $|\overline{\mathcal{M}_N^{(2)}}|^2$ term actually belongs to the next-to-next-to-leading order. However, if one invokes the chiral power counting to the amplitude,  up to $O(p^2)$ one should include all the three terms in Eqs.~\eqref{eq.mn1}-\eqref{eq.mn3}. In the top left panel of Fig.~\ref{fig.TCS_1}, we give the three different results by separately including the terms in Eqs.~\eqref{eq.mn1}-\eqref{eq.mn3}. The uncertainties for the full NLO amplitude squared in Fig.~\ref{fig.TCS_1}, i.e. the shaded zones with red color, are given by the parameters in Eq.~\eqref{eq.gapgan}. 
While for the neutron channel, the LO amplitude vanishes in the KSVZ case. For the predictions of the DFSZ case, we have used the full NLO $\gamma p \to a p$ amplitudes.

In Fig.~\ref{fig.TCS_2}, we show the curves by fixing $m_a=0$ for Scenario {\bf II} that only takes the contributions from the $t$-channel photon exchanges in the $\gamma N \to a N$ processes. In this scenario, the axion-photon-photon coupling $C_{a\gamma}$ enters the axion photoproduction amplitudes, instead of the axion-nucleon ones. For the KSVZ case, i.e., $E/N=0$, the NLO correction is almost negligible, comparing with the LO result. This occurs because the NLO anomalous magnetic moment term is strongly suppressed when the intermediate photon is highly off shell. The shaded uncertainty areas are estimated by the model-independent part of the $C_{a\gamma}$ as elaborated below Eq.~\eqref{eq.gapp}. Generally speaking, the cross sections contributed by the pure photon exchanges are around 2-3 orders smaller than those from the nucleon exchanges in the KSVZ case. For the DFSZ situation, the results are quite sensitive to the values of $E/N$.

In Fig.~\ref{fig.TCS_VM1}, the curves are shown by fixing $m_a=0$ for Scenario {\bf III} that only includes the contributions from the $t$-channel exchanges of the vector mesons in the $\gamma N\to a N$ processes. For the KSVZ case that is not affected by the unknown model-dependent axion parameters, the magnitudes of both cross sections of the proton and neutron processes contributed by the $t$-channel vector-meson exchanges are several orders larger than those in Scenario {\bf II}, which only includes the photon exchanges in the $t$ channel. The shaded uncertainty areas in Fig.~\ref{fig.TCS_VM1} are estimated via the $Va\gamma$ parameters in Eq.~\eqref{eq.vag1num} for Set-I and Eq.~\eqref{eq.vag2num} for Set-II. For the neutron process, the contributions from the vector-meson exchanges are the dominant ones over the nucleon exchanges; while for the proton process, the contributions from the vector-meson exchanges are around 1 order smaller than those from the nucleon exchanges. 
Regarding the DFSZ case, the resulting magnitudes of cross sections are at the same order as those from the KSVZ situation, by taking a large interval of $\sin^2\beta$. By including the vector-meson exchanges in different theoretical frameworks, the $\gamma p\to a p$ and $\gamma n \to a n$ cross sections are also calculated in Refs.~\cite{Aloni:2019ruo,Chakraborty:2024tyx}, respectively. We show their results in Fig.~\ref{fig.TCS_VM1} for comparisons. For the proton channel, there are some mild tensions between ours and the results from Ref.~\cite{Aloni:2019ruo}. The chiral EFT is expected to  provide reliable results in the low-energy region away from the baryon resonances. For the neutron channel, our results agree well with that from Ref.~\cite{Chakraborty:2024tyx}.  

\begin{figure}[tb]
\centering
\includegraphics[width=1\textwidth]{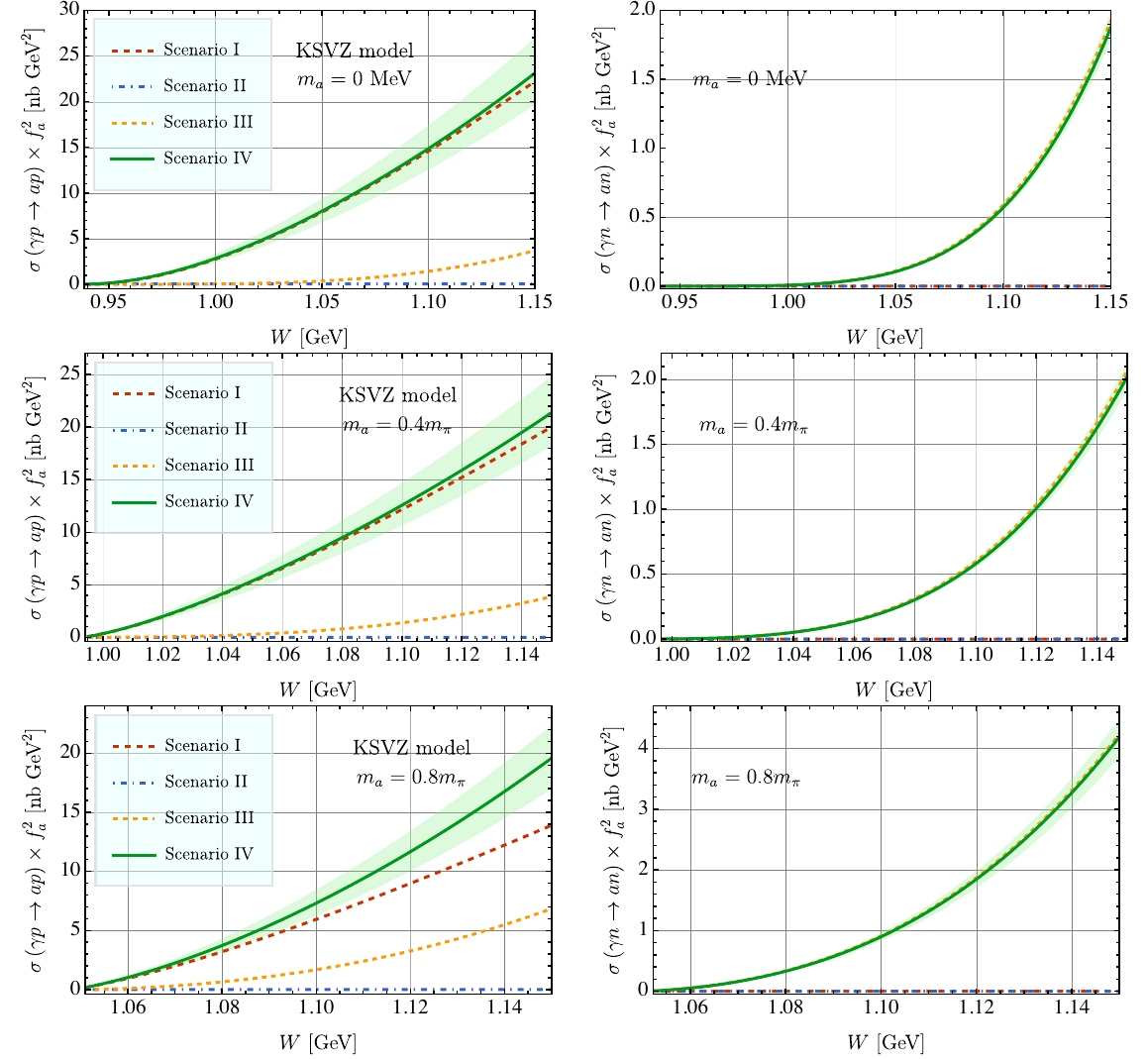} 
\caption{A comparison of the cross sections for the $\gamma p\to a p$ (left) and $\gamma n \to a n$ (right) processes between different scenarios of Eq.~\eqref{eq.csfour} in the KSVZ model, also known as the model-independent contributions to the axion amplitudes. Three different sets of axion masses, namely $m_a=0$, 0.4$m_\pi$ and 0.8$m_\pi$ are taken. The shaded zones correspond to the uncertainties, see the text for details. }\label{fig.TCS_tot}
\end{figure}

\begin{figure}[tb]
\centering
\includegraphics[width=1\textwidth]{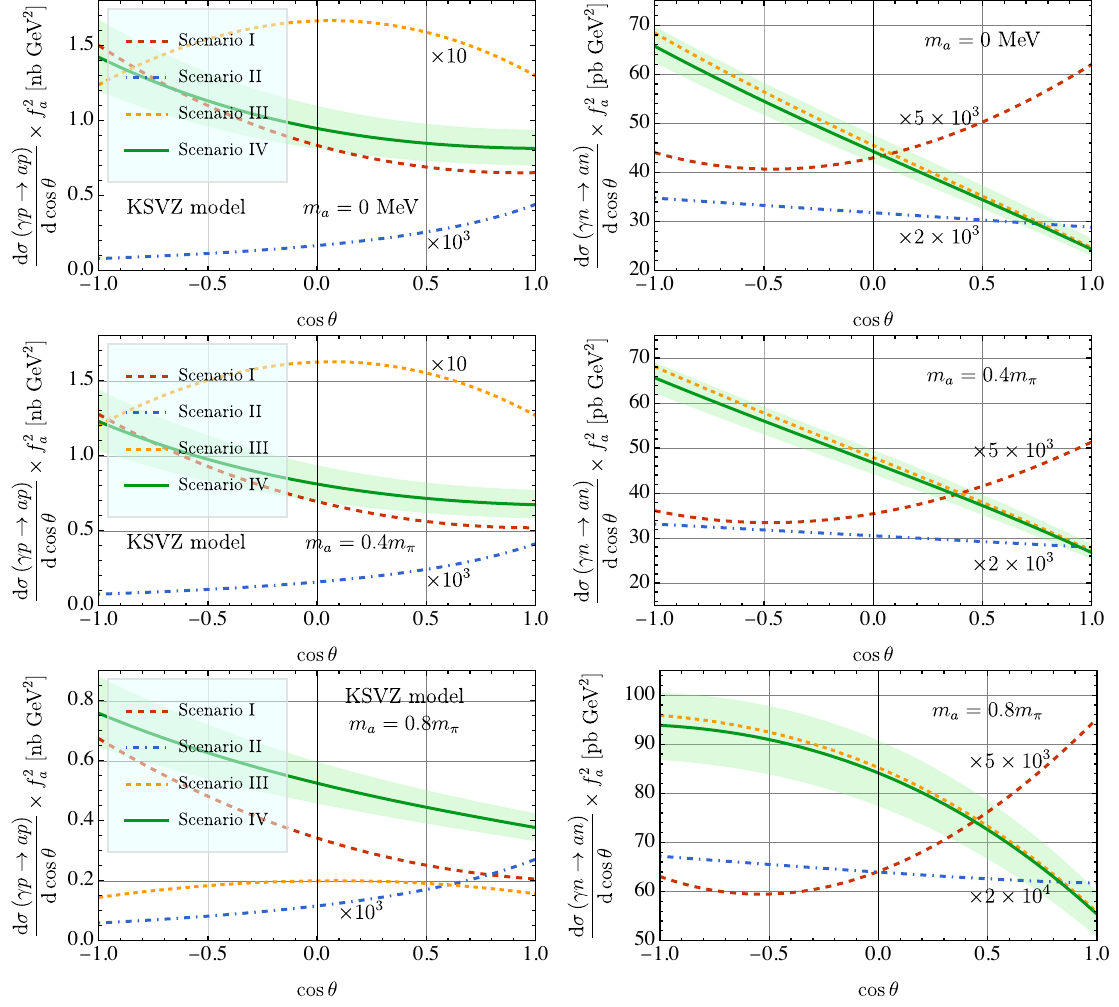} 
\caption{As Fig.~\ref{fig.TCS_tot} but for the differential cross section of Eq.~\eqref{eq.diffcs} at the c.m. energy $W=1.1~$GeV.}\label{fig.TCS_diffcs}
\end{figure}

In Fig.~\ref{fig.TCS_tot}, we summarize the results of the four scenarios in Eq.~\eqref{eq.csfour} for the KSVZ case by taking three different values of the axion masses, i.e., $m_a=0$, 0.4$m_\pi$ and 0.8$m_\pi$. The shaded uncertainty areas shown in Figs.~\ref{fig.TCS_tot} and \ref{fig.TCS_diffcs} are estimated by adding in quadrature all relevant parameters mentioned previously, including $g_{aN}$, $C_{a\gamma}$ and $g_{Va\gamma}$.  For the $\gamma p \to a p$ process, the dominant contributions come from the nucleon exchanges, as indicated by the close curves between Scenario I and Scenario IV in Fig.~\ref{fig.TCS_tot}, especially for the axion with small masses. While for the neutron process, the $t$-channel vector-meson exchanges overwhelmingly predominate over the other two mechanisms and the results are insensitive to the axion masses. 

Apart from the total cross section discussed above, we also study the differential cross sections
\begin{align}\label{eq.diffcs}
\frac{\md \sigma}{\md \cos\theta}=\frac{|\mathbf{p}(s)|}{|\mathbf{k}(s)|}\frac{|\overline{\mathcal{M}}|^2(s,t)}{32 \pi s}\ ,
\end{align}
where $\theta$ denotes the scattering angle in the c.m. frame and it is related to the Mandelstam variable $t$ through $t=m_a^2-\frac{(s-m_N^2)(s+m_a^2-mN^2)}{2s}+|\mathbf{k}(s)||\mathbf{p}(s)|\cos\theta$. In some cases, although the total cross sections show similar trends in different scenarios, significant variances in the differential cross sections could appear. Therefore, the differential cross sections can provide useful quantities to discriminate different mechanisms in the $\gamma N \to a N$ processes. For illustration, we give the differential cross sections at $W=1.1$~GeV for the KSVZ case by taking three different axion masses. According to the curves shown in Fig.~\ref{fig.TCS_diffcs}, the shapes of the distributions with respect to $\cos\theta$ arising from different mechanisms in Scenarios {\bf I}, {\bf II} and {\bf III} are obviously  distinctive, although the magnitudes differ significantly as well. Nevertheless, since the magnitudes could be influenced by the model-dependent axion parameters to a great extent, the shapes of the $\cos\theta$ distributions are expected to be insensitive to such inputs. Therefore the measurements of such differential distributions can be very helpful to distinguish different axion models.

\section{Summary and Conclusions}\label{sec.summary}

In this work we rely on the chiral effective field theory to perform a thorough study of the axion photoproduction off the nucleon, i.e. the $\gamma N\to a N$ processes, by simultaneously taking into account several typical mechanisms, instead of just focusing on one specific type of possible contributions. The nucleon exchanges in the $s$ and $u$ channels are calculated up to next-to-leading order. Not only effects from the $a\gamma\gamma$ interacting vertex, but also the contributions from the vector-meson exchanges, are incorporated in our calculation. Various pertinent  couplings in our study are fixed to a great extent by taking various hadronic inputs, such as the nucleon matrix elements of the axial-vector quark currents, the related $V\to P\gamma$ processes, the QCD short-distance constraints, etc. Meanwhile, both the KSVZ and DFSZ axion models are considered in the calculation. 
However we should also mention that some approximations have been used in our calculation. E.g., the $U(1)_A$ anomaly effect is ignored for the isosinglet current, and excited baryon states and higher order pieces of the amplitudes beyond $O(p^2)$ are not considered. The isospin breaking corrections are kept in the coefficients of the quark axial-vector currents involving axion, but they are neglected in the hadronic matrix elements. Although such neglected effects are not expected to significantly modify the current results, at least at a qualitative level, it would be interesting to pursue a more involved study by including them in a future work.

The simultaneous inclusion of different types of contributions allows us to carry out thoroughgoing analyses of the phenomenologies of the $\gamma N\to a N$ processes.  Different behaviors of the total and differential cross sections are revealed by separately including distinct microscopic mechanisms in the $\gamma p\to a p$ and $\gamma n\to a n$ processes. The axion photoproduction amplitudes calculated in this work are expected to provide reliable inputs for future studies of related processes.

\section*{Acknowledgements}
We thank Feng-Kun Guo and Hai-Qing Zhou for useful discussions and careful reading of the paper. XHC thanks Qian-Qian Guo for checking some numerical results. This work is funded in part by the National Natural Science Foundation of China (NSFC) under Grants Nos.~12150013, 11975090, 12347120 and 12475078; the Science Foundation of Hebei Normal University with contract No.~L2023B09; and the Postdoctoral Fellowship Program of China Postdoctoral Science Foundation under Grant Nos. GZC20232773 and 2023M74360.

%\newpage
% \bibliographystyle{unsrt}
\bibliography{ref}
\end{document}